\documentclass[twocolumn,aps,superscriptaddress,prl]{revtex4}
\usepackage{ifpdf}
\usepackage{graphicx}
\usepackage{times}
\usepackage{amsmath}
\usepackage{amssymb}
\usepackage{units}
\usepackage{stmaryrd} 

\renewcommand{\vec}[1]{\boldsymbol{#1}}

\begin{document}
\preprint{0}

\title{Anisotropic spin gaps in BiAg$_{2}$-Ag/Si$(111)$}

\author{Alberto Crepaldi}
email[alberto.crepaldi@epfl.ch]
\affiliation{Institut of Condensed Matter Physics
(ICMP), Ecole Polytechnique F\'ed\'erale de Lausanne (EPFL),
CH-1005 Lausanne, Switzerland}

\author{St\'ephane Pons}
email[stephane.pons@insp.jussieu.ch]
\affiliation{Institut of Condensed Matter Physics
(ICMP), Ecole Polytechnique F\'ed\'erale de Lausanne (EPFL),
CH-1005 Lausanne, Switzerland}
\affiliation{Institut des NanoSciences de Paris, CNRS : UMR7588, Universit\'e Pierre et Marie Curie - Paris VI, France}

\author{Emmanouil Frantzeskakis}
\thanks{Present address: Synchrotron SOLEIL, L'Orme des Merisiers, Saint Aubin-BP 48, 91192 Gif sur Yvette Cedex, France}
\affiliation{Institut of Condensed Matter Physics
(ICMP), Ecole Polytechnique F\'ed\'erale de Lausanne (EPFL),
CH-1005 Lausanne, Switzerland}

\author{Klaus Kern}
\affiliation{Institut of Condensed Matter Physics
(ICMP), Ecole Polytechnique F\'ed\'erale de Lausanne (EPFL),
CH-1005 Lausanne, Switzerland}
\affiliation{Max-Planck-Institut f\"ur Festk\"orperforschung, Heisenbergstrasse 1, 70569 Stuttgart, Germany}

\author{Marco Grioni}
\affiliation{Institut of Condensed Matter Physics
(ICMP), Ecole Polytechnique F\'ed\'erale de Lausanne (EPFL),
CH-1005 Lausanne, Switzerland}
\date{\today}

\begin{abstract}
We present a detailed analysis of the band structure of the $\sqrt{3} \times \sqrt{3}$ R $30^\circ$ BiAg$_2$/Ag/Si(111) trilayer system by means of high resolution Angle Resolved Photoemission Spectroscopy (ARPES). BiAg$_2$/Ag/Si(111) exhibits a complex spin polarized electronic structure due to giant spin-orbit interactions. We show  that a complete set of constant energy ARPES maps, supplemented by a modified nearly free electron calculation, provides a unique insight into the structure of the spin polarized bands and spin gaps. We also show that the complex gap structure can be continuously tuned in energy by a controlled deposition of an alkali metal.
\end{abstract}

\maketitle
\begin{figure*}
  \centering
  \includegraphics[width = 12 cm]{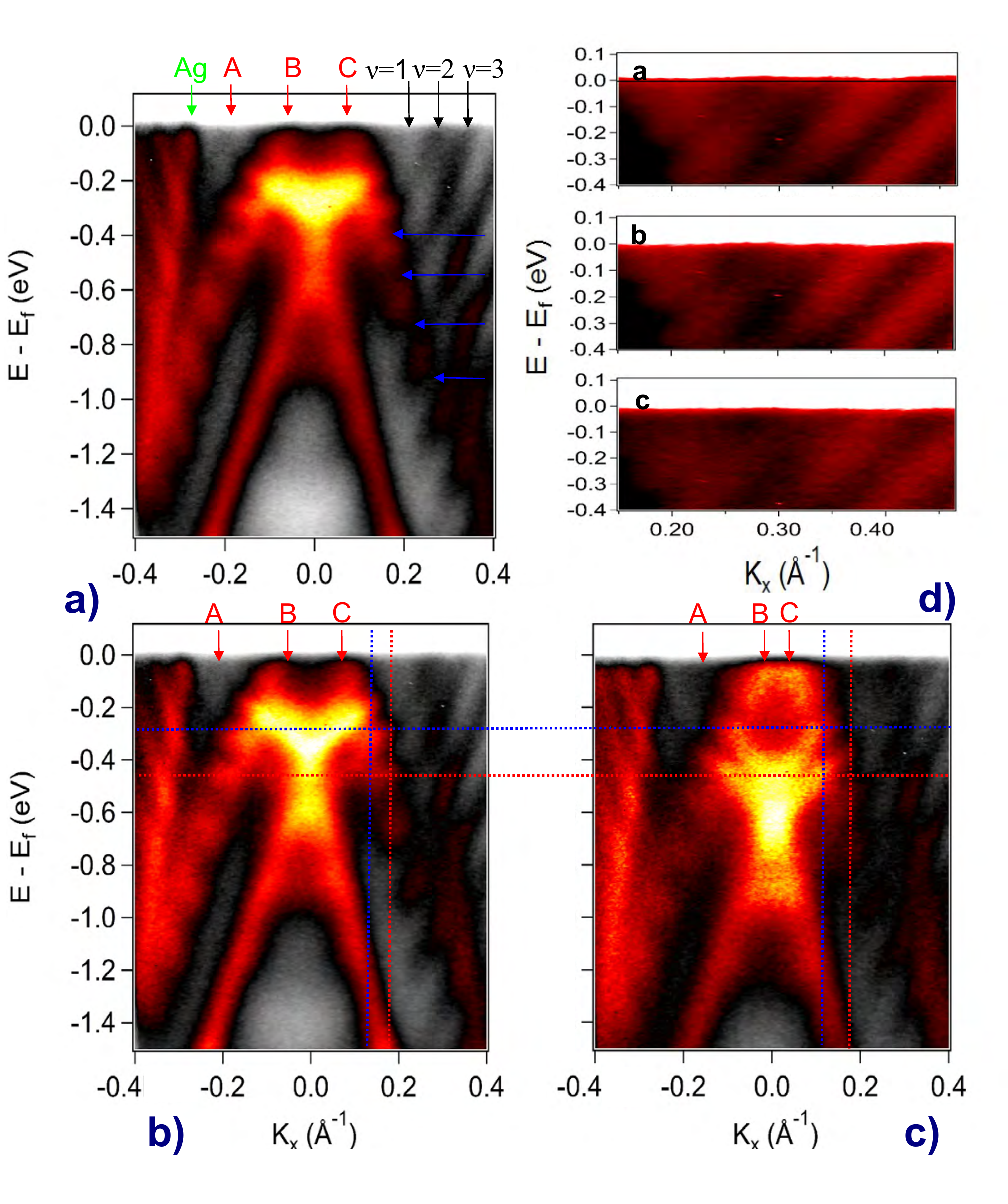}
  \caption{(a) ARPES intensity map of BiAg$_2$/Ag/Si(111) for $\theta$=19ML. The first three silver quantum well states $(\nu=i$ with $i=1,2,3)$ are indicated by black vertical arrows. The deposition of 1/3ML of Bi yields fully occupied $\left| {sp_z } \right\rangle $ derived surface states (SS), and $\left| {sp_{xy} } \right\rangle $ derived SS crossing E$_{F}$ (vertical red arrows A, B, C). Bands gap open at the crossing of the spin split
  states with the nearly degenerate QWS. A downward $\sim$ 100~meV
energy shift is observed for the Ag QWS after the formation of the surface alloy, consistent with previous reports
\cite{Hirada2008prb,frantz2008prl}. The feature marked bv the vertical arrow labelled Ag is due to scattering of photoelectrons emitted from a QWS on the periodic potential of the reconstruction.
  (b) and (c) were measured after two subsequent Na depositions.
   Red (blue) dashed lines in (b) cross at the
presence (absence) of a spin gap. After reaching the
saturation Na coverage, (c), the same
crossings are now associated with the absence (presence) of the
gap. (d) Corresponding close-ups of the QWS dispersion near E$_{F}$.}
  \label{fig:doping}
\end{figure*}

The control over the polarization of the electronic states close to the Fermi energy and the capability to transport spin polarized
curents represent two fundamental requirements for spintronic
applications \cite{zutic2004rmp,sun2005prb,koga2002prl}. The discovery of a giant Rashba effect in surface alloys has highlighted
the potential of spin split states at surfaces and interfaces \cite{ast2007prl} to achieve these goals.
In these systems, the net polarization of the density of states is null because of time reversal symmetry. Nevertheless, the Rashba interaction introduces a momentum-dependent effective magnetic field which separates the spin density of states in the reciprocal space~\cite{Rashba1984}. Such effective magnetic field and spin separation of the bands are expected to enable the control of spin polarized currents \cite{Datta1990,berciu}. Recent experiments indicate the fascinating prospect of transferring the spin split surface states to semiconductor
substrates \cite{frantz2008prl,he2008PRL,sakamoto2009prl,gierz2009prl,Hatta2009,yaji2010nature,Miyata2011}, bringing these systems one step closer to future
technological applications. In this work, we present a detailed analysis of the band structure
properties of the BiAg$_{2}$/Ag/Si(111) trilayer system.
We show that constant energy (CE) maps provide a
unique source of information on the topology of the spin polarized
surface states and gaps. We also show that the energy of the gap structure can be continuously tuned by a controlled
deposition of an alkali metal (Na).

A Si(111) crystal (n-doped, resistivity 0.011 $\Omega$cm) was cleaned
by repeated cycles of direct current heating, with flashes at $1400$K followed by steps at $900$K. The sample was slowly cooled across the 7$\times$7
transition temperature during the last
preparation cycle. The quality of the 7$\times$7 reconstruction was checked by means of
low-energy electron diffraction (LEED). Ag films (thickness $>$ 8 ML) were evaporated
from a tungsten basket onto the cold substrate ($100$K). After a mild post-annealing, Ag 1$\times$1 spots appeared in the LEED pattern. This protocol yields a layer-by-layer growth with a sharp incommensurate interface \cite{Speer2006}.
The deposition of $1/3$ ML Bi was achieved with an EFM3 Omicron evaporator. After a post-annealing at RT, $\sqrt{3} \times \sqrt{3} $ R $30^\circ$ extra spots in the LEED pattern indicated the formation of a BiAg$_2$ surface alloy. Na was evaporated from a commercial dispenser (SAES getter) onto the cold sample ($100$K).

Angle-resolved photoelectron spectroscopy (ARPES) was performed at
$70$K with a Phoibos 150 Specs analyzer. The energy resolution
was set equal to 10 meV. The partially polarized UV light (HeI$\alpha$ line at $21.22$ eV) was produced by a monochromatized high
brightness Gammadata VUV 5000 lamp.

\begin{figure*}
  \centering
  \includegraphics[width = 14 cm]{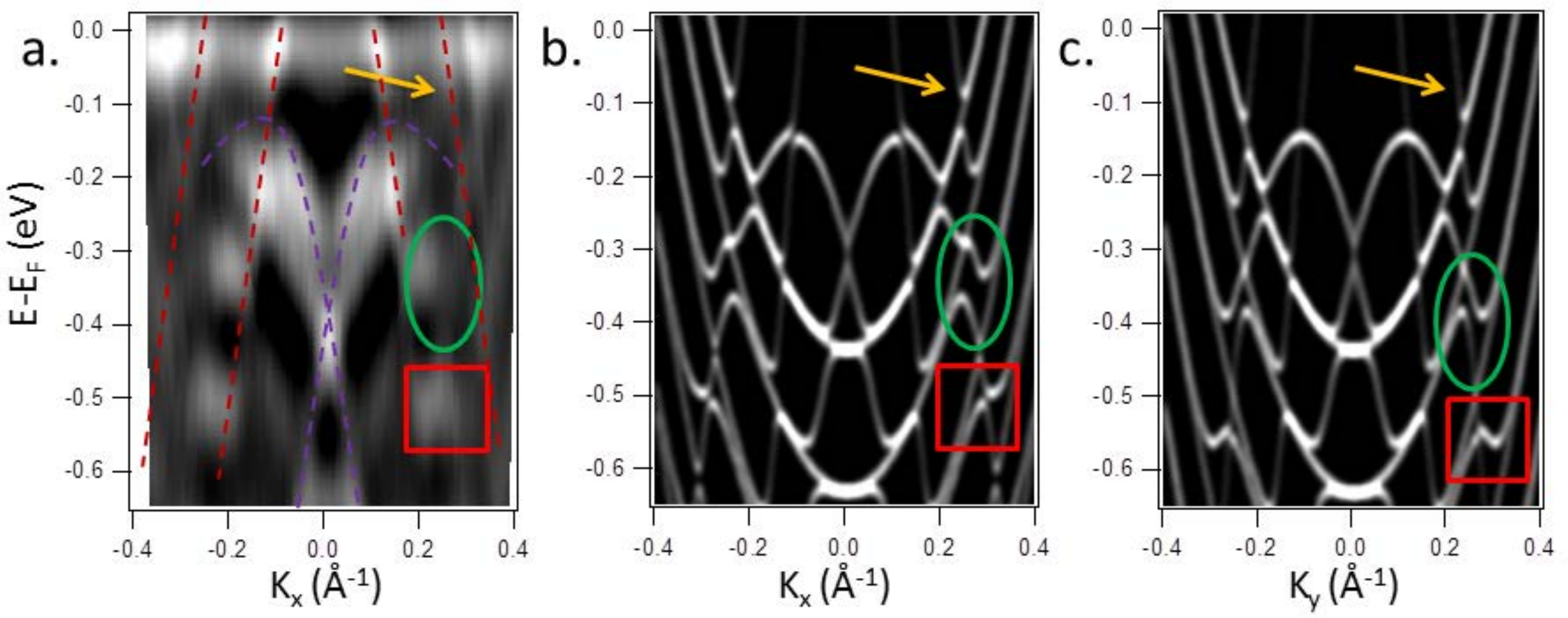}
  \caption[didascalia]{(color online) a) Derivative of the photoemission intensity along the $\bar \Gamma \bar K$ direction (k$_x$) for a silver thickness of 25ML. Dashed lines are guide to the eyes and indicate the surface states. The silver quantum well states are localized in the subsurface planes so they have a low spectral weight with respect to the surface states. b) and c) Calculated band structure around $\bar\Gamma$ along the $\bar \Gamma \bar K$ (a) and $\bar \Gamma \bar M$ (b) high symmetry directions. The calculation parameters are given in the appendix. The QWS positions agree well with a silver thickness of 25ML. White stands for high intensity. The broadening and the intensity of the calculated bands are the result of an artificial gaussian convolution with an arbitrary width. The arrow and the boxes point out some peculiar patterns described in the text.}
  \label{fig:kxky}
\end{figure*}

The interplay of a large spin-orbit splitting and of quantum confinement in the BiAg$_2$/Ag/Si(111) system leads to a complex electronic structure, presented Fig.~\ref{fig:kxky}(a) and Fig.~\ref{fig:doping} (a) \cite{frantz2008prl,sakamoto2009prl,franz2010jes,he2010prl}. Its origin can be summarized as follows:
\begin{description}\setlength{\leftskip}{-0pt}\item[a.]\noindent Quantum well states (QWS) develop in the silver layer due to the confinement of
the Ag s-states by the vacuum on one side, and by the fundamental energy gap on the substrate side \cite{Speer2006}.  The
QWS have a nearly-free electron character, and give rise to a set of parabolic sub-bands with small and positive effective masses around $\bar \Gamma$, the center of the Brillouin zone (BZ) Fig.~\ref{fig:doping} (a). The number and energy positions of these sub-bands are determined by the thin film thickness. A standard Rashba mechanism \cite{Rashba1984} splits the QWS, but the splitting is too small to be directly observable \cite{frantz2008prl,he2008PRL}.
\item[b.]\noindent  A downward $\sim$ 100~meV energy shift is observed for the Ag QWS after the formation of the surface alloy, consistent with previous reports\cite{Hirada2008prb,frantz2008prl}. Two-dimensional SS with negative effective masses develop in the BiAg$_2$ surface alloy. They have mainly $\left| {sp_z } \right\rangle $ and, respectively, $\left| {sp_{xy} } \right\rangle $ character. Each band is split into two strongly spin polarized sub-bands as a result of a strong Rashba interaction, similar to what was previously found in BiAg$_2$/Ag(111) \cite{ast2007prl}. The $\left| {sp_z } \right\rangle $ states are split in momentum by $\Delta k_{sp_z}=0.13$~ {\AA}$^{-1}$, and cross at $\bar \Gamma$ at a binding energy $E_{B}^{sp_z}=350$~meV. The two $\left| {sp_{xy} } \right\rangle $ spin components cross $\sim500$ meV above the Fermi level (E$_F$) \cite{bihl2007prb, ast2008prb}. All the SS bands of BiAg$_2$ deviate from a simple parabolic dispersion and show a hexagonal anisotropy \cite{ast2007prl,Premper2007}.
\item[c.]\noindent Spin conservation \cite{Didiot2006} and symmetry considerations determine the hybridization of the spin polarized SS bands with the nearly degenerate quantum well states. As a result the BiAg$_2$/Ag/Si(111) system exhibits a spin polarized and multi-gapped electronic structure \cite{frantz2008prl,he2008PRL,franz2010jes,he2010prl}.
\end{description}

The ARPES intensity map of Fig.~\ref{fig:doping} (a) illustrates the dispersion of the hybridized $\left| {sp } \right\rangle $ SS and $\left| {s } \right\rangle $ QWS. Spin gaps can already be identified in the image. The high contrast enables a precise analysis of the intensity distribution, namely close to E$_F$ and around the gaps. We show in the following that the hexagonal anisotropy of the SS dispersion generates a peculiar spectral density pattern in this quantum-confined Rashba system.

\begin{figure*}
  \centering
  \includegraphics[width = 18 cm]{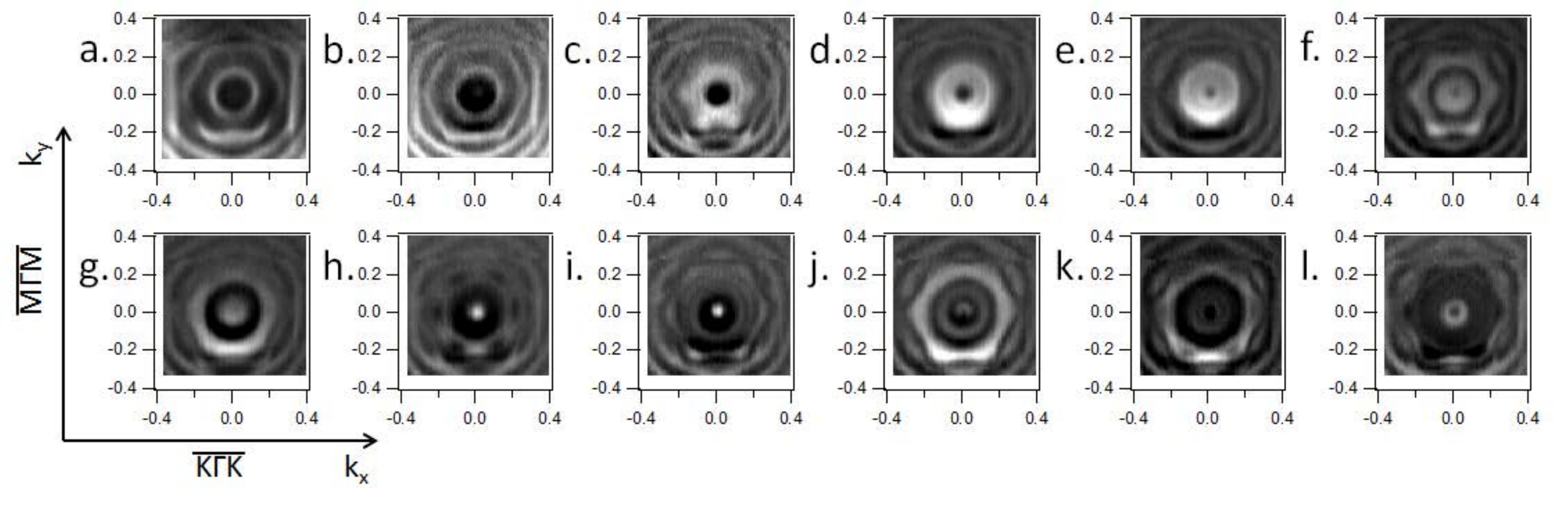}
  \caption{Second derivative images of the constant energy ARPES intensity maps close to $\bar\Gamma$ for the following binding energies:  a. $50$meV, b. $100$meV, c. $150$meV, d. $200$meV, e. $250$meV, f. $300$meV, g. $350$meV, h. $400$meV, i. $450$meV, j. $500$meV, k. $550$meV, l. $600$ meV. $k_x$ and $k_y$ vary from -0.4 to +0.4 {\AA}$^{-1}$ in each frame. the sample corresponds to a silver thickness of 25ML. White stands for high intensity. The spots at the center of the images are an artifact of the derivative procedure. }
  \label{fig:CE1}
\end{figure*}

We restrict our ARPES study to a region of reciprocal space around $\bar \Gamma$ covering the first Brillouin Zone of the $\sqrt{3} \times \sqrt{3} $ R $30^\circ$ reconstruction. Fig.~\ref{fig:CE1} presents a set of constant energy cuts through the band structure, for several binding energies between 50 meV and 600 meV.
The two contours closest to $\bar\Gamma$ in panel (a) correspond to the spin polarized $\left|sp_{xy}\right\rangle$ SS of the BiAg$_2$ alloy  \cite{frantz2008prl}. These bands exhibit specific features of the giant Rashba splitting: the inner contour is circular, while the outer one is hexagonal. The hexagonal third contour around $\bar\Gamma$ is the backfolded signal of an Ag QWS. It is due to the scattering of the photoelectrons from the $\sqrt{3} \times \sqrt{3}$ R $30^\circ$ surface potential, and therefore is a final state effect of the ARPES process. This feature partly masks the circular contours of the Ag QWS in Fig.~\ref{fig:CE1} (a) and (b).
The signatures of the Ag QWS are visible at larger momenta, near the corners of the image. At higher binding energy, in panel (c), the additional hexagonal contour at $\vec k\approx(0.2,0.2)$\AA$^{-1}$ reflects the top of the spin-orbit split $\left|sp_z\right\rangle$ bands of the alloy. With increasing binding energy the contours of the QWS shrink, while those of the SS expand, according to the opposite sign of their effective masses.

A description of the band structure is hampered by the large number of bands and by the hexagonal anisotropy of the multigapped structure.  \emph{Ab-initio} calculations of the constant energy contours face the difficulties of a complex interface structure, and of the dense sampling of $k$ space. We chose instead a phenomenological approach, based on an anisotropic nearly free electron (NFE) model, with additional terms describing the hybridization of the surface and quantum wells bands.

Due to the lack of structural inversion symmetry of the (crystal, interface, vacuum) system, both the SS and the QWS are expected to exhibit a Rashba behavior \cite{Rashba1984} and to be at least partially polarized. A spin polarization of the QWS, which has been observed e.g. for Pb thin films on Si(111) \cite{Dil_PRL_Pb}, is a prerequisite for the opening of spin gaps \cite{frantz2008prl,he2008PRL}. However, the Rashba coupling is so weak that the splitting cannot be directly observed, and the Ag QWS appear to be degenerate and parabolic.  The sixfold symmetry of the SS contours is the result of the threefold rotational symmetry of the alloy structure and time reversal symmetry. The giant SO splitting of the bands has been attributed to i) an anisotropy in the in-plane surface potential \cite{ast2007prl,Premper2007}, ii) the buckling of the Bi alloy layer, contributing to the anisotropy of the SS wavefunction along the surface normal \cite{Bentmann2009}; or again iii) the unquenching of the orbital momentum at the surface \cite{Park2011}. Time reversal symmetry requires that the two spin branches of a Rashba system have opposite polarization, so that each branch can be considered separately in discussing their hybridization with other states.
Recently, the snow-flake-like contour of the Fermi surface of Bi$_2$Te$_3$ was simulated by adding higher order terms to an effective Hamiltonian \cite{Fu2009,Basak2011}. A similar NFE calculation for the Bi/Si(111) interface \cite{Frantzeskakis2010PRB} yields a good agreement with the experiment \cite{EmmanouilThesis,Frantzeskakis2011}.
We extend that approach to the present case. We used the following spin Hamiltonian with the basis states $\left| {\vec k, \uparrow } \right\rangle $, $\left| {\vec k, \downarrow } \right\rangle $, ($\vec k^2=k_x^2+k_y^2$), to describe the spin polarized branches of the $\left| {sp_z } \right\rangle $ and $\left| {sp_{xy} } \right\rangle $ SS:
\begin{widetext}
\begin{equation}
{H_{{\rm{sp}}}}(\vec k) = \left( {\begin{array}{*{20}{c}}
{E_0^{{\rm{sp}}} + \frac{1}{2}{c_{{\rm{sp}}}}({{({k_x} - {\rm{i}}{k_y})}^3} + {{({k_x} + {\rm{i}}{k_y})}^3}) + \frac{\vec k^2}{{2{m_{{\rm{sp}}}}}}}&{{\alpha _{{\rm{sp}}}}( - {\rm{i}}{k_x} - {k_y})}\\
{{\alpha _{{\rm{sp}}}}({\rm{i}}{k_x} - {k_y})}&{E_0^{{\rm{sp}}} + \frac{1}{2}{c_{{\rm{sp}}}}({{({k_x} - {\rm{i}}{k_y})}^3} + {{({k_x} + {\rm{i}}{k_y})}^3}) + \frac{\vec k^2}{{2{m_{{\rm{sp}}}}}}}
\end{array}} \right) \end{equation}
\end{widetext}
where $m_{sp}$, $E_0^{sp}$, $\alpha_{sp}$ are respectively the negative effective mass of the SS, the binding energy of the surface band and the effective Rashba constant, and $c_{sp}$ is an anisotropy parameter. The parameters for the $\left| {sp_z } \right\rangle $ and $\left| {sp_{xy} } \right\rangle $ states are different.
Each spin-branch of the QWS is described by a parabolic band of $\left| {s} \right\rangle $ character:
\begin{widetext}
\begin{equation}
{H_{{\rm{QWS},\nu}}}(\vec{k}) =\left( {\begin{array}{*{20}{c}}
{E_{0,\nu}^{{\rm{QWS}}} + \frac{{\vec k^2}}{{2{m_{{\rm{QWS}}}}}}}&0\\
0&{E_{0,\nu}^{{\rm{QWS}}} + \frac{{\vec k^2}}{{2{m_{{\rm{QWS}}}}}}}
\end{array}} \right)\end{equation}
\end{widetext}
where the effective mass of the QWS, $m_{QWS}$, is the same whatever the index ($\nu$) of the quantum well band.
$E_{0,\nu}^{QWS}$, the binding energy of the $\nu^{th}$ quantum well band, follows the standard confinement law along one dimension. $E_{0,\nu}^{QWS}$ and $m_{QWS}$ can be fit to the unperturbed QWS dispersion far from the SS.
The parameters $V_\nu$ describing the hybridization between the quantum well states $\left| {QWS,\nu,\sigma} \right\rangle $ and the $\left| {sp_z,\sigma' } \right\rangle $ and $\left| {sp_{xy},\sigma' } \right\rangle $ SS were adapted from \cite{he2008PRL}, and assumed independent of the orbital character of the SS. A full polarization for the QWS is enforced by setting the hybridization to $0$ for opposite spin states; i.e.
\begin{equation}
\left\langle {QWS,n,\sigma } \right|H\left| {sp,\sigma '} \right\rangle = V_\nu \; \rm{if}\; \sigma=\sigma' ~,
\end{equation}
and
\begin{equation}
\left\langle {QWS,n,\sigma } \right|H\left| {sp,\sigma '} \right\rangle = 0 \; \rm{if}\; \sigma\ne\sigma' ~.
\label{hyb}
\end{equation}
As suggested by group theory considerations \cite{Mirhosseini2009,Vajna2011} and \emph{ab-initio} calculations~\cite{ast2007prl,frantz2008prl,bihl2007prb}, we also introduce a weak coupling between the $\;\left| {sp_z } \right\rangle $ and $\;\left| {sp_{xy} } \right\rangle $ SS:
\begin{equation}
\left\langle {sp_z} \right|H\left| {sp_xy} \right\rangle = V~.
\end{equation}
A weak but nonzero coupling is necessary to achieve a good agreement with the data. This additional term would lead to a contradiction with eq.~(\ref{hyb}) for $V$ large with respect to $V_\nu$, which is not the case here.
All parameters (see appendix) were determined from the best fits to the data of Fig.~\ref{fig:CE1}. The QWS parameters correspond to a thickness of around 25 ML \cite{Matsudaa2002}.
Despite its simplicity, the model captures the main features of the experimental data. The hexagonal anisotropy of the surfaces states is not perfectly reproduced near the top of the bands. The inner bands are closer to $\bar \Gamma$ and more circular then in the experimental maps. The agreement improves further away from $\bar \Gamma$. The model does not include the backfolding of the SS.

Fig.~\ref{fig:kxky}(b) and (c) show the simulated band dispersion in the $k_x$ ($\bar \Gamma \bar K$) and, respectively, $k_y$ ($\bar \Gamma \bar M$) directions of the reconstructed BZ (parameters correspond to a thickness of around 25 ML, see appendix). It should be compared to Fig.~~\ref{fig:kxky}a) acquired for the same thickness. The experimental data show a stronger spectral weight for the SS than for the QWS. Indeed, the surface states are strongly localized at the topmost layer whereas the QWS lie deeper in the Ag layer. The z-delocalization of the states is not computed so the balance of intensity between the SS and the QWS cannot be reproduced. The QWS are unperturbed at large $k$ values as in the experimental data.
The spin gaps appear here as diamond shapes, e.g. in the (green) oval box. The NFE model predicts that the diamond structures are strongly spin polarized, in agreement with \emph{ab-initio} calculations~\cite{frantz2008prl}. Because of the hexagonal symmetry of the SS, the avoided crossings with the QWS -- i.e. the gaps -- occur at different energies in different directions, e.g. in the (red) rectangular box, and the width of the diamond structures is also different.

\begin{figure*}
  \centering
  \includegraphics[width = 18 cm]{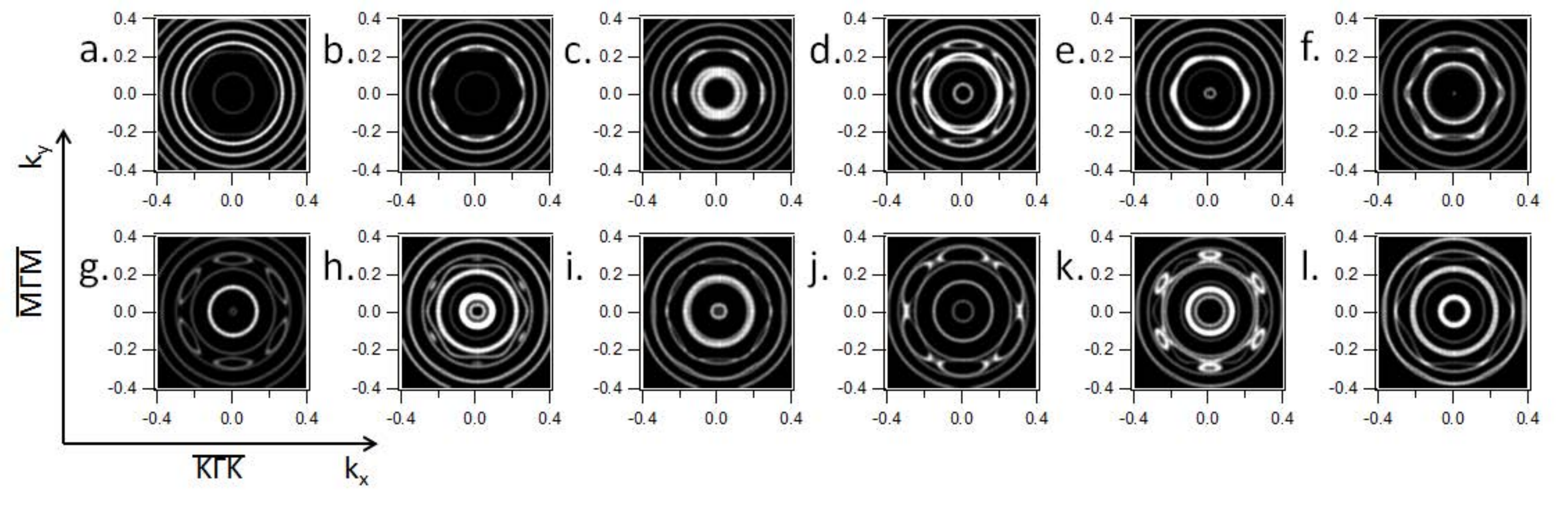}
  \caption[didascalia]{Calculated CE contours close to $\bar\Gamma$ for the following binding energies:  a. $50$meV, b. $100$meV, c. $150$meV, d. $200$meV, e. $250$meV, f. $300$meV, g. $350$meV, h. $400$meV, i. $450$meV, j. $500$meV, k. $550$meV, l. $600$ meV. $k_x$ and $k_y$ vary from -0.4 to +0.4 {\AA}$^{-1}$ in each frame. The calculation parameters are given in the appendix. The QWS positions agree well with a silver thickness of 25ML. White stands for high intensity. The broadening and the intensity of the calculated bands are the result of an artificial gaussian convolution with an arbitrary width.}
  \label{fig:CE2}
\end{figure*}

A comparison of the simulated constant energy map and the calculated band diagram of Fig.~\ref{fig:kxky} yields further insight into the complex gap structure. As expected, only the $\left| {sp_{xy} } \right\rangle $ and the QWS contribute to the spectral intensity near E$_F$ (Fig.~\ref{fig:CE2} (a)). In Fig.~\ref{fig:CE2} (b) (E$_B$=$100$ meV) the hybridization between the $\left| {sp_{xy} } \right\rangle $ SS and the first QWS induces a small decrease of spectral intensity at $k_x \approx 0.3${\AA}$^{-1}$ and at the five equivalent wave vectors rotated by $60^\circ$ from each other. In Fig.~\ref{fig:CE2} (c) (E$_B$=$150$ meV; on the other side of the energy gap) the spectral weight is largest at the same BZ locations. Such a constrast inversion is observed whenever the $\left| {sp_{xy} } \right\rangle $ or $\left| {sp_{z} } \right\rangle $ SS cross a QWS. The arrows in Fig.~\ref{fig:kxky} (b) and (c) show that it reflects a hybridization gap in one of the spin polarized component of the QWS. Thus, the QWS acquire a hexagonal anisotropy close to the gaps.

Further from E$_F$ the band structure is more complex, because the sequence of the bands is affected by hybridization. Fig.~\ref{fig:CE2} (d) and (e) (E$_B \approx 200-250$ meV) show an interesting energy range where the outer $\left| {sp_{xy} }  \right\rangle $ band hybridizes with the $1^{st}$ QWS, and the outer $\left| {sp_{z} } \right\rangle $ band hybridizes with the $2^{nd}$ QWS. In this window of energy, only one spin branch of each of the two first QWS contributes to the band diagram in the $\bar \Gamma \bar K$ direction whereas there is no spin gap in the $\bar \Gamma \bar M$ direction (this is also visible in Fig.~\ref{fig:kxky}). The resulting band topology yields strongly anisotropic polarization patterns.
The two circular signatures close to the center of the image correspond to the top of the $\left| {sp_{z} } \right\rangle $ bands.
In Fig.~\ref{fig:CE2} (j) (E$_B$=$500$ meV) closed contours are visible at $k_y \approx$0.3{\AA}$^{-1}$ and at the equivalent wave vectors. They correspond to wiggles in the dispersion, inside the (red) square box of Fig.~\ref{fig:kxky}, where an NFE model predicts strongly spin polarized states (see e.g. \cite{he2008PRL}). By contrast, unpolarized states are expected at $k_x \approx$~0.3~{\AA}$^{-1}$ and at the equivalent wave vectors, where two bands with opposite polarization cross.

The difference in the $\bar \Gamma \bar M $ and $\bar \Gamma \bar K$ band structure is expected to introduce a different response for the electrons involved in the conduction properties along two orthogonal directions. The viability of a future device based on a similar Rashba system
depends on the possibility of adjusting the energy positions of the spin gaps relative to the Fermi level.
We have already shown that the electronic structure of
the trilayer can be controlled by varying the thickness of the Ag layer
\cite{frantz2008prl}. However, only discrete shifts
in the energy position of the quantum well states could be obtained, corresponding to
discrete changes of the boundary conditions for the confined $\left| s \right\rangle $ Ag states.
This falls short of the precise control of the spin gap texture required for spintronics applications. Here we
explore a different strategy, and dope the trilayer system by depositing an alkali metal (Na).
We show below that this approach enables a fine tuning of the band structure.

The ARPES intensity maps of Fig.~\ref{fig:doping} illustrate the effect of doping. Fig.~\ref{fig:doping} (a) is a reference for the BiAg$_2$/Ag/Si(111) surface. The initial deposition of a small amount of Na (panel (b)) induces a downward energy shift ($-50$ meV) of the $\left| {sp_{z} } \right\rangle $
SS. The energy shift increases with the amount of Na, and eventually saturates (panel (c)) at a coverage $\theta \sim0.25$ ML, estimated by comparison with the case of Na on Bi/Cu(111) \cite{Bentmann2009}.

At the saturation, the total energy shift of the two fully occupied $\left| {sp_{z} } \right\rangle $ SS is $\sim-250$ meV, similar to what was observed for BiAg(111) \cite{franz2010jes}. Besides, at saturation, the total shift ($\Delta k_F$) of the inner $\left| {sp_{xy} } \right\rangle$ branches is equal to $0.05${\AA }$^{-1}$. While the inner branches of the $\left| {sp_{xy} } \right\rangle$ derived SS ($k_{F}$=$\pm0.08$~{\AA }$^{-1}$ before doping, and $k_{F}$=$\pm0.03$~{\AA }$^{-1}$ at saturation) are clearly resolved (red arrows B and C in Fig.~\ref{fig:doping}). The outer branches, which cross E$_F$ at $k_{F}=\pm0.21${\AA}$^{-1}$ \cite{ast2007prl,franz2010jes}, are much weaker due to ARPES matrix elements, and a Fermi level crossing can be identified only for negative $k$ values (arrow A in Fig.~\ref{fig:doping}).
Figure~\ref{fig:doping} (d) presents a close-up of the QWS
dispersion for positive $k$ values, before (a) and after (b, c) two Na depositions cycles.
Contrary to the Bi-derived SS,
the energy of the QWS is not affected by the Na adsorption. Even at saturation
there is no change in the corresponding Fermi wave vectors.
Clearly, the electrons donated by the Na atoms dope the BiAg$_{2}$ SS, which are strongly localized at the surface, rather than the deeper lying QWS. The result is a continuous shift of the SS with respect to the QWS as a function of Na coverage, and a corresponding
shift of the gap structure relative to the Fermi level. The partially spin polarized gap structure is so strongly affected by the alkali doping that the position of the gap (indicated by the crossing of the blue dashed lines in Fig.~\ref{fig:doping} (b)) can be swept with the fully un-gapped region (blue lines crossing in Fig.~\ref{fig:doping} (c)) and viceversa (crossing of the red lines in Fig.~\ref{fig:doping} (b) and (c)). Similar effect, even if not experimentally resolved, must affect also the spin gap close to the Fermi level.

In summary, ARPES reveals the topology of the
spin-dependent hybridization gaps of the trilayer BiAg$_{2}$/Ag/Si(111)
system which exhibits a giant Rashba effect. Doping of the surface states by the deposition of controlled amounts of sodium
opens the way to a complete control of the energy position of the band
structure. The ability of tuning the Fermi level across states with strong and opposite polarization
suggests possible future applications in spintronics.
We expect that our novel
experimental data will stimulate further theoretical work
on the spin-dependent transport properties of Rashba surface alloys.

\section{Acknowledgements}
E.F. acknowledges the Alexander S. Onassis Public Benefit Foundation for the award of a scholarship. This research was supported by the Swiss NSF and the NCCR MaNEP. S.P. thanks Tristan Cren for his help with technical aspects of the calculations.

\section{Appendix}
\begin{tabular}{|l|r|}
\hline
\multicolumn{2}{|c|}{NFE calculation parameters} \\
\hline
$E_0^{sp_{xy}}$ & $500$ meV \\
$m_{sp_{xy}}$ & $-0.019$ \\
$\alpha_{sp_{xy}}$ & $2.9$ \\
$c_{sp_{xy}}$ & $36$ \\
\hline
$E_0^{sp_{z}}$ & $-350$ meV \\
$m_{sp_{z}} $ & $-0.035$ \\
$\alpha_{sp_{z}}$ & $3.1$ \\
$c_{sp_{z}}$ & $19$ \\
\hline
$E_{0,1}^{QWS}$ & $-430$ meV \\
$E_{0,2}^{QWS}$ & $-620$ meV \\
$E_{0,3}^{QWS}$ & $-860$ meV \\
$E_{0,4}^{QWS}$ & $-1150$ meV \\
$E_{0,5}^{QWS}$ & $-1400$ meV \\
$m_\nu$ & $0.092$ \\
\hline
$V$ & $20$ meV \\
$V_1$ & $25$ meV \\
$V_2$ & $50$ meV \\
$V_3$ & $70$ meV \\
$V_4$ & $100$ meV \\
$V_5$ & $120$ meV \\
\hline
\end{tabular}

\end{document}